\documentclass[useAMS,usenatbib]{mn2e}
\usepackage{epsfig}
\newcommand{\beq}{\begin{equation}}
\newcommand{\eeq}{\end{equation}}

\newdimen\hssize
\newdimen\hdsize 
\title[Gas depletion in cluster galaxies]
{Gas depletion in cluster galaxies depends strongly on their internal structure}
\author[Zhang et  al.]{Wei Zhang$^{1}$\thanks{E-mail:xtwfn@bao.ac.cn},   
  Cheng   Li$^{2}$, Guinevere Kauffmann$^{3}$, Ting Xiao$^{2}
  \thanks{LAMOST fellow}$
  \\          
  ${^1}$  National  Astronomical
  Observatories,  Chinese Academy of  Sciences, 20A Datun Road, Chaoyang District, Beijing  100012, China
  \\  ${^2}$  
  Partner Group of Max Planck Institut for Astrophysics and Key Laboratory 
  for Research in Galaxies and Cosmology 
  \\
  ~~~of Chinese Academy of Sciences, Shanghai Astronomical Observatory,  
  Nandan Road 80, Shanghai 200030, China       
  \\          
  ${^3}$     Max    Planck    Institut     f\"ur    Astrophysik,
  Karl-Schwarzschild-Strasse  1,  85748  Garching, Germany  
}             
              
\begin{document}
              
              
\date{Accepted ........ Received ........; in original form ........}
              
\pagerange{\pageref{firstpage}--\pageref{lastpage}} \pubyear{2012}
              
\maketitle    
              
\label{firstpage}
              
\begin{abstract}
We analyze galaxies in 300 nearby groups and clusters identified in the Sloan
Digital Sky Survey using a  
photometric gas mass indicator that is useful for estimating the degree to
which the interstellar medium of a cluster galaxy has been depleted. 
We  study the radial
dependence of inferred gas mass fractions for galaxies of different stellar masses and stellar
surface densities. 
At fixed clustercentric distance 
and at fixed stellar mass, lower density galaxies are more strongly depleted of their gas
than higher density galaxies.  
An analysis of depletion trends in the two-dimensional plane of
stellar mass $M_*$  and stellar mass surface density $\mu_*$ reveals that
gas depletion at fixed clustercentric radius is much more sensitive to
the density of a galaxy than to its mass. 
We suggest that low density
galaxies are more easily depleted of their gas, because
they are more easily affected by ram-pressure and/or tidal forces. 
We also look at the dependence of our  gas fraction/radius relations on the velocity dispersion of
the cluster, finding no clear systematic trend. 
\end{abstract}

\begin{keywords}
galaxies: clusters:  general --  galaxies: distances and  redshifts.
\end{keywords}

\section{Introduction}

The observed variation in galaxy colours, star formation rates, cold gas
fractions  as a function of distance from the centers of groups and
clusters places important constraints on the physical processes that affect
the gas in  galaxies as they evolve within a hierarchy of merging dark
matter halos \citep[e.g.][]{Diaferio-01,Okamoto-Nagashima-03}.

Semi-analytic models of galaxy formation developed in the early 1990's
\citep[e.g.][]{Kauffmann-White-Guiderdoni-93,Cole-94} made the  simplistic
assumption that after a galaxy was accreted by a larger dark matter halo
and became a satellite, its reservoir of hot gas would be stripped
instantaneously and would form part of the hot atmosphere bound to the
common halo. It was also assumed that the cold gas in a galaxy was not
affected by stripping processes and the timescale for the satellite galaxy
to redden was set by the rate at which the cold gas was used up  by star
formation.  Comparison of these models with group and cluster galaxies from
the Sloan Digital Sky Survey revealed that the predicted fraction of blue
satellite galaxies was too low \citep{Weinmann-06}. This discovery was
then followed by considerable effort to fix the models by relaxing the
assumption that the hot gas reservoir around a satellite is stripped
instantaneously following accretion \citep[e.g.][]{Font-08,Weinmann-10,Guo-11}.
If satellite galaxies are able to retain a
significant fraction of their hot gas for several gigayears following
accretion, their colours are found to be in much better agreement with
data.

One important constraint from the observations that has not been considered
in much detail up to now, is the fact that the increase in the fraction of
red galaxies from the outskirts of rich groups and clusters  to their
centers, is strongly dependent on galaxy mass.  As shown in Figure 7 of \citet{vonderLinden-10}
, the fraction of red galaxies with stellar masses
in the range $3 \times 10^9 M_{\odot}$ to $10^{10} M_{\odot}$ increases by a
factor 4 from  0.2 at the virial radius of the cluster to 0.8 at
the cluster center. In contrast, for galaxies with stellar masses greater
than $5 \times 10^{10} M_{\odot}$, the red fraction only increases by 30\%
from 0.6 at the virial radius to 0.8 at the cluster center. Because cold
gas consumption times in low mass field galaxies are long, one might ask
whether stripping of an  external reservoir of ionized or hot gas can lead
to such strong effects with cluster-centric radius. In a recent paper, \citet{Guo-11}
implemented a model in which the hot gas mass around
satellites is reduced in direct proportion to the mass of its surrounding
dark matter subhalo, which loses mass continuously due to tidal stripping.
In addition, Guo et al computed the radius at which ram-pressure forces due
to the satellite's motion through the intracluster medium would remove the
hot gas. The mimimum of the tidal radius and the ram-pressure stripping
radius define the radius beyond which gas is removed from the subhalo.  As
shown in their Figure 2, this model does not reproduce the stellar mass
dependence of cluster galaxy red fractions at  distances less than
$\sim 0.5 R_{vir}$ from the cluster center.

In this paper, we analyze 300 clusters and groups from the samples of
\citet{Berlind-06} and \citet{vonderLinden-07}, looking for clues
as to why environmental effects on low mass galaxies are so dramatic. In
section 2, we describe the cluster and galaxy samples and introduce a
photometric gas mass indicator that is useful for estimating the degree to
which the interstellar medium of a cluster galaxy has been depleted with
respect to a similar galaxy in the field. In section 3, we study the radial
dependence galaxy colours/gas fractions in the two-dimensional plane of
stellar mass $M_*$  and stellar mass surface density $\mu_*$. At fixed
cluster-centric radius, we find that decrease in inferred cold gas mass
fraction is a stronger function of stellar surface mass density than
stellar mass.  
In section 4, we  look at the
dependence of stripping effects on the cluster/group velocity dispersion
and in section 5, we summarize and discuss the implications of our results.
Throughout this paper we assume a spatially flat concordance cosmology
with $\Omega_m=0.3$, $\Omega_\Lambda=0.7$ and $H_0=100h$kms$^{-1}$Mpc$^{-1}$,
where $h=0.7$.

\section{Data} 

\begin{figure}
  \begin{center}
    \epsfig{figure=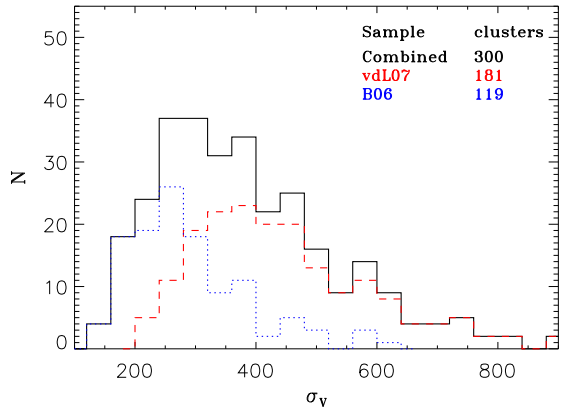,width=0.48\textwidth}
  \end{center}
  \caption{Histograms of the velocity dispersion $\sigma_V$ of the clusters
in our sample. The black solid histogram is for the full sample, while the
red dashed and blue dotted histograms are for the subsets from \citet{vonderLinden-07}
and \citet{Berlind-06}.} 
  \label{fig:sigmaV} 
\end{figure}

\begin{figure}
  \begin{center}
    \epsfig{figure=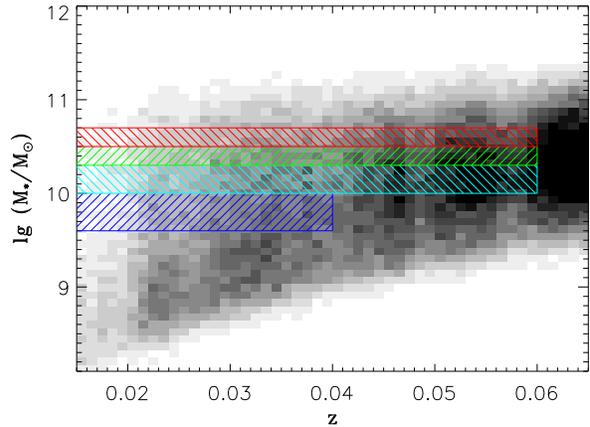,width=0.48\textwidth}
  \end{center}
  \caption{ The shaded contours show the distribution of 
parent sample galaxies in the plane of stellar mass versus redshift.
The four coloured boxes indicate the redshifts limits of the
four stellar mass sub-samples analyzed in this paper.} 
  \label{fig:mstar_z} 
\end{figure}

\subsection{The cluster catalogue}

The cluster catalogue used in this paper includes 300 unique galaxy clusters,
of which 181 were identified by \citet[][hereafter vdL07]{vonderLinden-07} 
from the SDSS data release 4 (DR4) and 119 were identified by 
\citet[][hereafter B06]{Berlind-06} from an earlier SDSS data release 
(DR3). Our cluster sample is  restricted to have redshift $z<0.06$ (see
next section), to have richness 
$N_{member}\ge10$, velocity dispersion $\sigma_V \ge 150$kms$^{-1}$ and
area completeness $f_{cover} \ge 0.5$. The area completeness is defined by
\begin{equation}
  f_{cover}=\frac{A_{survey}(<2r_{200})}{A_{full}(<2r_{200})}, 
\end{equation}
where $A_{full}(<2r_{200})$ is a circular sky area centered on the brightest
cluster galaxy (BCG) with  radius twice the virial radius of the cluster
($r_{200}$), and $A_{survey}$ is the area inside the same circle that is 
covered by the SDSS survey. The area completeness is  unity for clusters 
far enough from the survey edges, but can be very low for clusters  near the edges. 
Following \cite{Finn-05}, we estimate a virial radius for each of our clusters, 
$r_{200}$ as
\begin{equation}
\rm r_{200}=1.73\frac{\sigma_v}{1000km~s^{-1}}
\frac{1}{\sqrt{\Omega_\Lambda+\Omega_m(1+z)^3}}h^{-1}Mpc. 
\end{equation}

In Figure~\ref{fig:sigmaV} we show a  histogram of $\sigma_V$ for the 
cluster sample.  
We note that the vdL07 sample is
actually a subset of clusters in the C4 catalogue of \citet{Miller-05}, who
identified clusters based on the presence of a ``red sequence'' of galaxies
with similar positions and redshifts. vdL07 implemented an improved method for
identifying the brightest cluster galaxy (BCG) and recalculated the 
velocity dispersion for each cluster accordingly. The cluster-finding algorithm
of B06 was based on a redshift-space friends-of-friends method with no
requirement that there be a clearly defined red sequence.    
As can be seen from the figure, the clusters from vdL07 have higher velocity
dispersions than the clusters from B06. 
The combination of the two gives a sample with a wide coverage of
velocity dispersion, ranging from $\sim150$km/s to $\sim800$km/s. 
This corresponds to a wide range in dark matter halo mass, from 
$M_h\sim2\times 10^{12}M_\odot$ for the Milky Way-type halos 
up to $M_h\sim5\times 10^{14}M_\odot$ for the most massive halos in 
the local Universe \citep[see e.g.][]{Li-12c}.

In this paper, we show results only for the combined sample of clusters. 
We have tested that all conclusions presented in this paper hold
when we analyze the vdL07 and B06 cluster samples separately.
The quantitative relations between H{\sc i} gas mass fraction and
cluster-centric radius in bins of stellar mass and stellar
surface density agree for the two samples  within the statistical errors.
Combining the samples allows us to test whether there are any
clear differences in galaxy properties at fixed $r/r_{200}$
for  clusters with low and high velocity dispersions.
As we will show in Section 4, these differences are very small.

\begin{table}
  \caption{Volume-limited samples selected by stellar mass from the
    SDSS/DR7 galaxy sample}
  \begin{center}
    \begin{tabular}{cccc}\hline\hline
      Sample & Stellar mass & Redshift & $N_{gal}$   \\ \hline
      M1     & $9.60<\log(M_\ast/M_\odot)<10.0$ & $z<0.04$ & 4169 \\
      M2     & $10.0<\log(M_\ast/M_\odot)<10.3$ & $z<0.06$ & 9004 \\
      M3     & $10.3<\log(M_\ast/M_\odot)<10.5$ & $z<0.06$ & 5577 \\
      M4     & $10.5<\log(M_\ast/M_\odot)<10.7$ & $z<0.06$ & 4763 \\
      \hline
    \end{tabular}
  \end{center}   
  \label{tbl:samples}
\end{table}

\subsection{The galaxy sample}

We begin with the  parent galaxy sample  constructed from  the New York
University Value Added Catalogue (NYU-VAGC) {\tt sample dr72}
\citep{Blanton-05a}, which consists of about half a million galaxies  with
$r<17.6$, $-24<M_{^{0.1}r}<-16$ and redshifts in  the  range $0.01<z<0.5$.
Here,  $r$  is  the $r$-band  Petrosian apparent   magnitude,   corrected   for
Galactic   extinction,   and $M_{^{0.1}r}$ is the  $r$-band Petrosian absolute
magnitude, corrected for  evolution  and $K$-corrected  to  its  value  at
$z=0.1$.  We use stellar masses from the MPA/JHU SDSS/DR7 database
\footnote{http://www.mpa-garching.mpg.de/SDSS/DR7/}, which are estimated from
fits to the SDSS {\it ugriz} photometry following the philosophy of
\cite{Kauffmann-03a} and \cite{Salim-07}, assuming a universal stellar initial
mass function of \citet{Kroupa-01}.

From the parent sample, we select four volume-limited samples of galaxies
according to their stellar mass and redshift. The stellar mass range,
the redshift range and the number of galaxies of the samples are listed 
in Table~\ref{tbl:samples}. 
The four volume-limited samples include a total of 23,513 galaxies.
In Figure~\ref{fig:mstar_z} we indicate the
selection criteria of our samples in the stellar mass versus redshift
plane.  The parent sample is plotted in the background, grey-coded by
the number of galaxies. The samples we choose  are the same as in 
\citet[][hereafter vdL10; see the left-hand panel of their fig.5]
{vonderLinden-10}, except that we impose an upper
redshift cut at $z=0.06$.

\subsection{H{\sc i} mass fraction and H{\sc i} deficiency}

When discussing the effect of the cluster environment on galaxies, it is
much more physically intuitive to carry out analyses using 
{\em cold gas mass fractions} 
rather than  colours or star formation rates. This is because
we are trying to understand the impact of processes such as ram-pressure
or tidal stripping on the interstellar medium of galaxies \citep{Kormendy-Bender-12}. 

In recent  years,
`pseudo' H{\sc i} gas mass estimates have been introduced that rely on   
the fact that the H{\sc i} gas mass fraction is strongly correlated with
properties such optical and optical/IR colours 
\citep[e.g.][]{Kannappan-04}. Subsequent studies  
established that a combination of colour and stellar surface mass density provides a more
accurate estimation of the H{\sc i} mass fraction 
\citep[e.g.][]{Zhang-09, Catinella-10, Li-12b}.

The most recent version of such estimators was proposed by
\citet[][hereafter L12] {Li-12b} and utilizes a  combination of four
galaxy parameters:
\begin{eqnarray}\label{eqn:hiplane}
  \log(M_{\mbox{H{\sc i}}}/M_\ast) & = & -0.325\log\mu_\ast-0.237(NUV-r) \nonumber \\
                     &   & -0.354\log M_\ast-0.513\Delta_{g-i}+6.504,
\end{eqnarray}
$M_{*}$ is the stellar mass;
$\mu_\ast$ is the surface stellar mass density
given by $\log\mu_\ast=\log M_\ast-\log(2\pi R_{50}^2)$ 
($R_{50}$ is the radius enclosing half
the total $z$-band Petrosian flux and is in units of kpc).  $NUV-r$ is the global
near-ultraviolet (NUV) to $r$-band colour. The $NUV$ magnitude is
provided by the GALEX pipeline and the $NUV-r$ colour is corrected for Galactic
extinction following \cite{Wyder-07} with $A_{NUV-r} = 1.9807A_r$ , where $A_r$
is the extinction in r-band derived from the dust maps of
\cite{Schlegel-Finkbeiner-Davis-98}.
$\Delta_{g-i}$ is the colour gradient defined as the difference
in $g-i$ colour between the outer and inner regions of the galaxy. The inner
region is defined to be the region within $R_{50}$ and the outer region
is the region between $R_{50}$ and $R_{90}$.
The estimator has been calibrated using samples of nearby galaxies ($0.025<z<0.05$)
with H{\sc i} line detections from the GALEX Arecibo SDSS Survey
\citep[GASS;][]{Catinella-10}, and is demonstrated to provide
unbiased H{\sc i}-to-stellar mass ratio estimates, $M_{\mbox{H\sc i}}/M_\ast$,
even for H{\sc i}-rich galaxies.

As well as the H{\sc i} mass fraction,  we will also work with an  `H{\sc i}
deficiency parameter', $H_{def}$ which we define as the deviation in
$\log(M_{HI}/M_\ast)$ from the value predicted from the mean relation between
$\log(M_{HI}/M_\ast)$ and galaxy mass $M_\ast$ and stellar surface mass density
$\mu_\ast$ \citep[see][]{Li-12b}: \begin{equation}\label{eqn:hidef}
H_{def}=\log(M_{\mbox{H{\sc i}}}/M_\ast)-\log(M_{\mbox{H{\sc
i}}}/M_\ast)|(M_\ast,\mu_*) \end{equation} where $\log(M_{\mbox{H{\sc
i}}}/M_\ast)$ is estimated by equation~(\ref{eqn:hiplane}) and
\begin{equation}\label{eqn:hiplane2} \log(M_{\mbox{H{\sc
i}}}/M_\ast)|(M_\ast,\mu_*)= -0.227\log M_\ast - 0.646 \log \mu_*+7.166.
\end{equation} The relation between the stellar mass of a galaxy and its
structural parameters such as stellar surface mass density and concentration
index have been found to depend extremely weakly on environment
\citep[e.g.][]{Kauffmann-04,Weinmann-09}.  The H{\sc i} deficiency parameter is
thus the most direct measure of the degree to which gas has been depleted in a
given cluster galaxy.

Because the photometric estimator in equation (\ref{eqn:hiplane}) has been
calibrated using the GASS sample, one might question whether it is still valid
for galaxies in cluster environments. \citet{Cortese-11} used a sample of
$\sim300$ nearby galaxies to investigate the effect of the environment on H{\sc
i} scaling relations and found that Virgo cluster galaxies still lie on the
same `plane' relating H{\sc i} gas mass fraction to stellar surface mass
density and colour, even though they are significantly offset towards lower gas
content compared to field galaxies.

\begin{figure*}
  \begin{center}
    \epsfig{figure=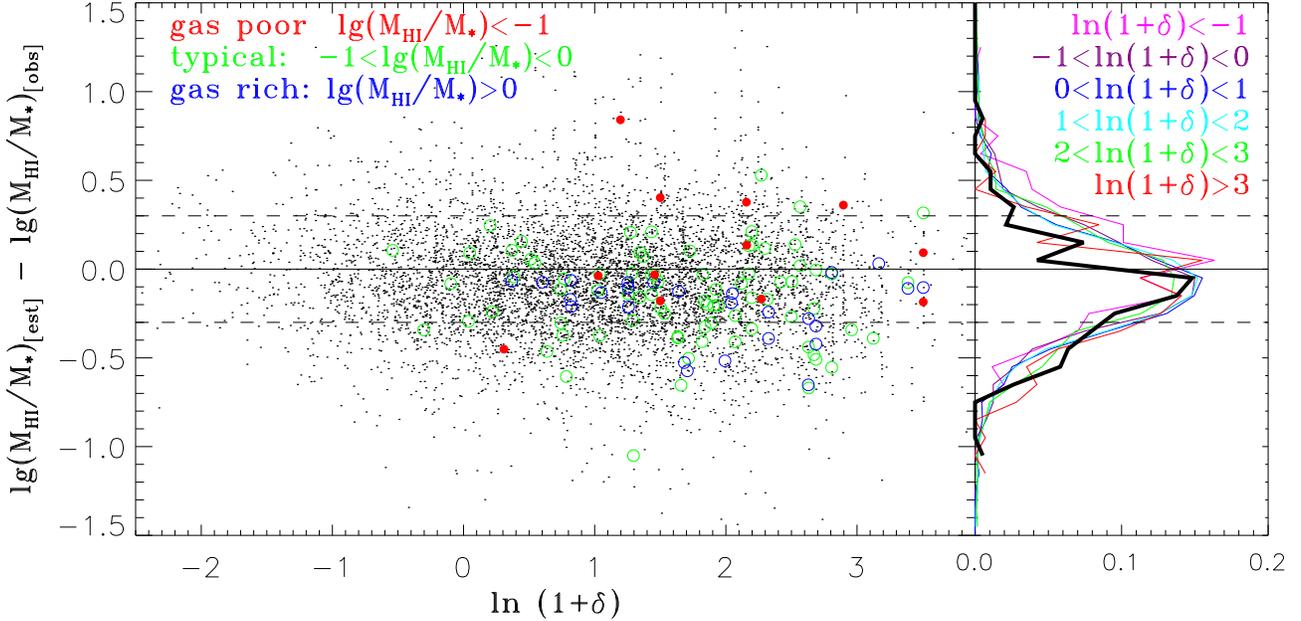,width=\textwidth}
  \end{center}
\caption{The residual
in the estimated H{\sc i} mass fraction, defined as the difference between
the estimated $\log({M_{HI}/M_\ast})$ value and the observed value, is plotted as
a function of overdensity, $\ln(1+\delta)$. In the left panel, 
the black points are SDSS galaxies detected by ALFALFA. 
The red, green and blue circles are group/cluster galaxies identified by \citet{Berlind-06}
that have been detected by ALFALFA. Red symbols show ``gas poor''
cluster galaxies, green symbols show ``typical'' cluster galaxies  and blue
symbols indicate  ``gas rich'' cluster galaxies, with H{\sc i} mass fractions as indicated
on the plot. In the right panel, the ALFALFA galaxies have been separated 
into six subsamples according to local density, as indicated on top of the panel. 
The distributions of gas fraction residuals of these subsamples are shown as  coloured lines.
The distribution of gas fraction residuals of the B06 cluster galaxies 
is shown as a thick solid black line.}
\label{fig:residual}
\end{figure*}

In Figure~\ref{fig:residual} we use  $\sim8000$ galaxies from the Arecibo
Legacy Fast ALFA survey\citep[ALFALFA;][]{Giovanelli-05} to test whether L12
estimator exhibits any significant environmental dependence. We plot the
residual in the estimated H{\sc i} mass fraction, defined as the difference
between the estimated $\log({M_{HI}/M_\ast)}$ value and the observed value, as
a function of overdensity, $\ln(1+\delta)$.  The overdensity parameter,
$\delta$ has been  estimated by \citet{Jasche-10} through reconstruction of the
3D density field from the SDSS DR7 data.  In the right-hand panel of the same
figure we plot the histograms of the residual for subsets of galaxies selected
by $\ln(1+\delta)$. We find  that the residual in the estimated H{\sc i} mass
fraction exhibits a weak, but systematic trend with overdensity, in the sense
that the gas fractions of galaxies in high-density regions are slightly
underestimated.  Cluster galaxies detected by ALFALFA that are included in the
B06 group/cluster catalogue are plotted as coloured symbols in
Figure~\ref{fig:residual} and the distribution of residuals is shown as a solid
black line in the right hand panel. As can be seen, the  H{\sc i} content of
cluster galaxies is underestimated by 0.1 dex on average, but the galaxies that
cause this shift are the {\em gas-rich} rather than the  gas-poor group/cluster
members. This particular paper is focused on the physical processes that cause
galaxies to become gas-deficient, so we will leave aside  this curious
phenomenon for the moment.    

\section{Results}

\begin{figure*} \begin{center} \epsfig{figure=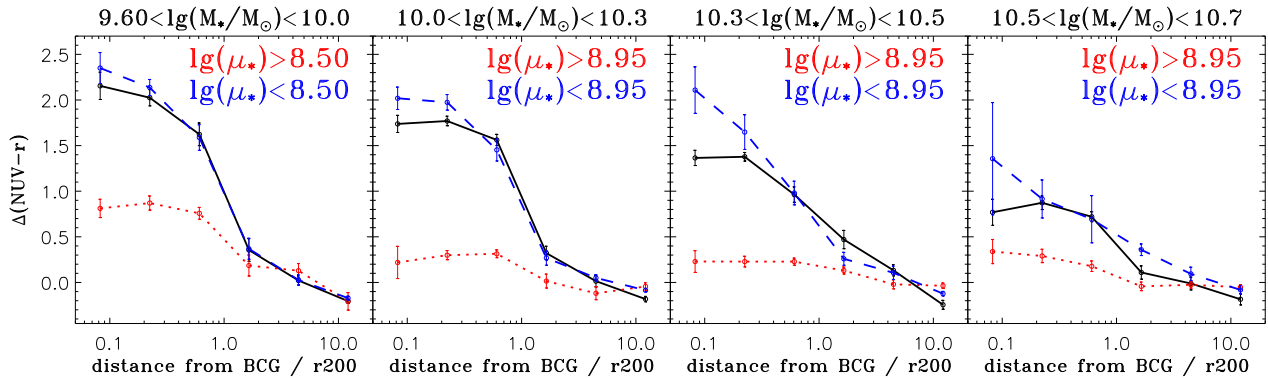,width=\textwidth}
\end{center} \caption{ The difference in the median  $NUV-r$ colour compared to
galaxies of the same stellar mass and stellar surface density in the field,
where the value of the ``field galaxies'' is given by the median value of the
subsample of the NYU-VAGC {\tt sample dr72} in the same coverage of stellar
mass and stellar surface density.  Results are plotted as a function of
distance from the BCG, scaled to the virial radius of the cluster.  The four
panels show results  for galaxies in different stellar mass intervals. Black
curves show results for all galaxies.  Galaxies with higher and lower surface
mass density $\mu_*$ are showed as red dotted and blue dashed lines (the
division points are labelled in each panel).  Errors are estimated via
bootstrap resampling.} \label{fig:nuv_radius_mustar} \end{figure*}

We begin by analyzing trends in the $NUV-r$ colours of galaxies as a 
function of cluster-centric radius. This is a directly observed (as opposed to inferred) 
quantity that has been found to correlate strongly with the H{\sc i} gas mass
fraction of nearby galaxies \citep{Catinella-10}. In 
Figure~\ref{fig:nuv_radius_mustar}, we plot the  {\em difference}
in the median  $NUV-r$ colour compared to galaxies of the same stellar mass  
and stellar surface density in the field, where the value of
the ``field galaxies'' is given by the median value of the
subsample of the NYU-VAGC {\tt sample dr72} in the same coverage of stellar mass and stellar surface density. Results are plotted 
as a function of distance from the BCG, scaled to the virial radius of the cluster.
The four  panels show results  for galaxies in different stellar 
mass intervals. 
\footnote {For a given stellar mass subsample we use the clusters below the same 
redshift limit of the galaxy subsample. As a result, for the lowest mass 
range, $9.6<\log_{10}(M_\ast/M_\odot)<10.0$, only the clusters with 
$z<0.04$ is used, while for the other three mass ranges all the clusters 
with $z<0.06$ are used.} 
Black lines show results for all galaxies
in a given mass interval, while red and blue curves show
results for  high and low values of $\mu_\ast$, respectively. 
Errors are estimated using a standard bootstrap resampling technique.

In all stellar mass intervals, the difference in $NUV-r$ color increases with
decreasing cluster-centric distance. The strength of the effect depends
strongly on stellar mass. For galaxies with stellar masses less than $10^{10}
M_{\odot}$, the $NUV-r$ colour reddens  by more than 2 magnitudes from the
virial radius to the center of the cluster. For galaxies with stellar masses
greater than $3 \times 10^{10} M_{\odot}$, the reddening is only $\sim 0.5$
magnitudes.  The very large differences between the blue and red curves in
Figure~\ref{fig:nuv_radius_mustar} show that the effect at fixed mass is mainly
driven by the dependence of the reddening  on surface mass density.  We will
come back to this point in more detail later.

\begin{figure*}
  \begin{center}
    \epsfig{figure=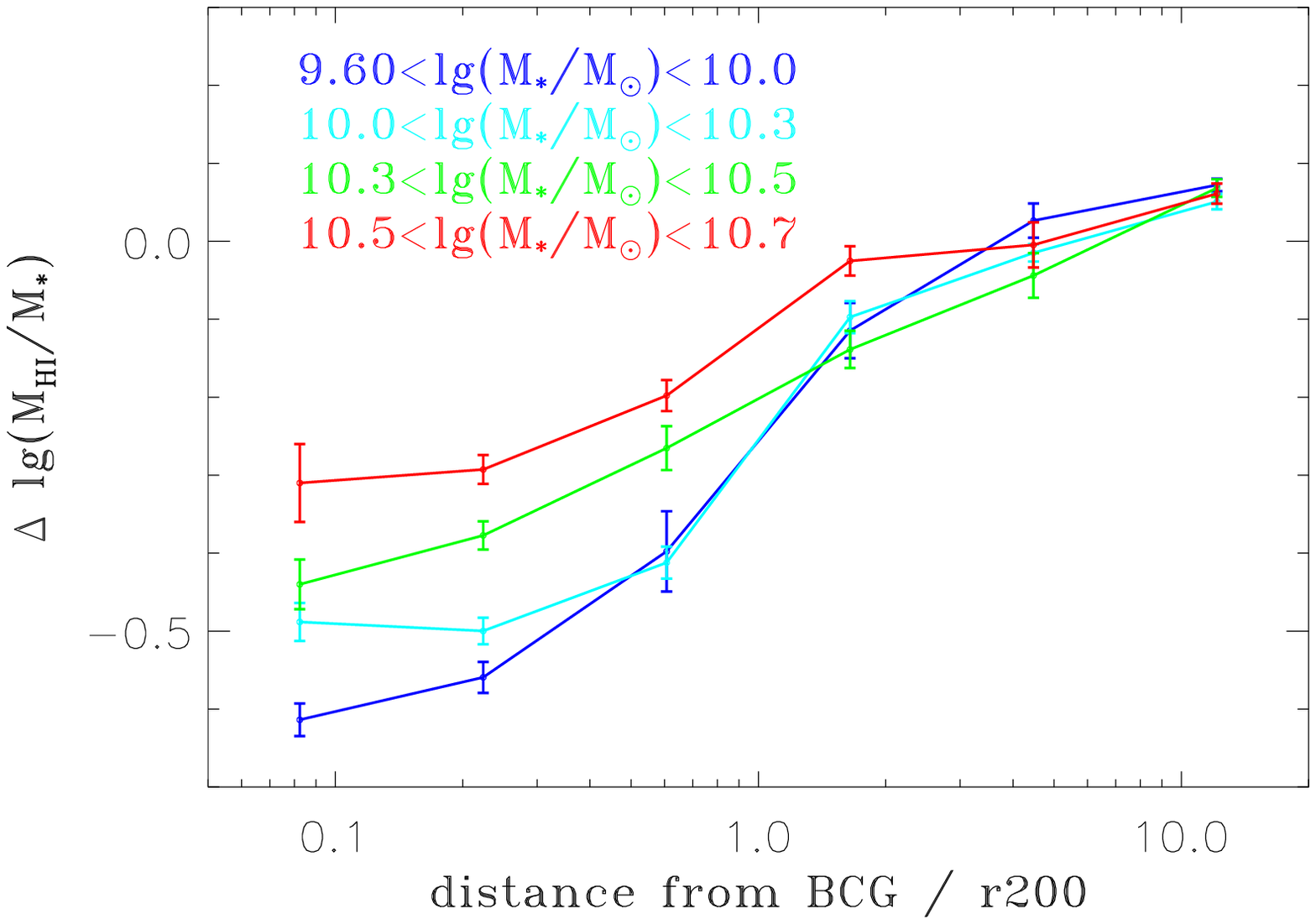,width=0.48\textwidth}
    \epsfig{figure=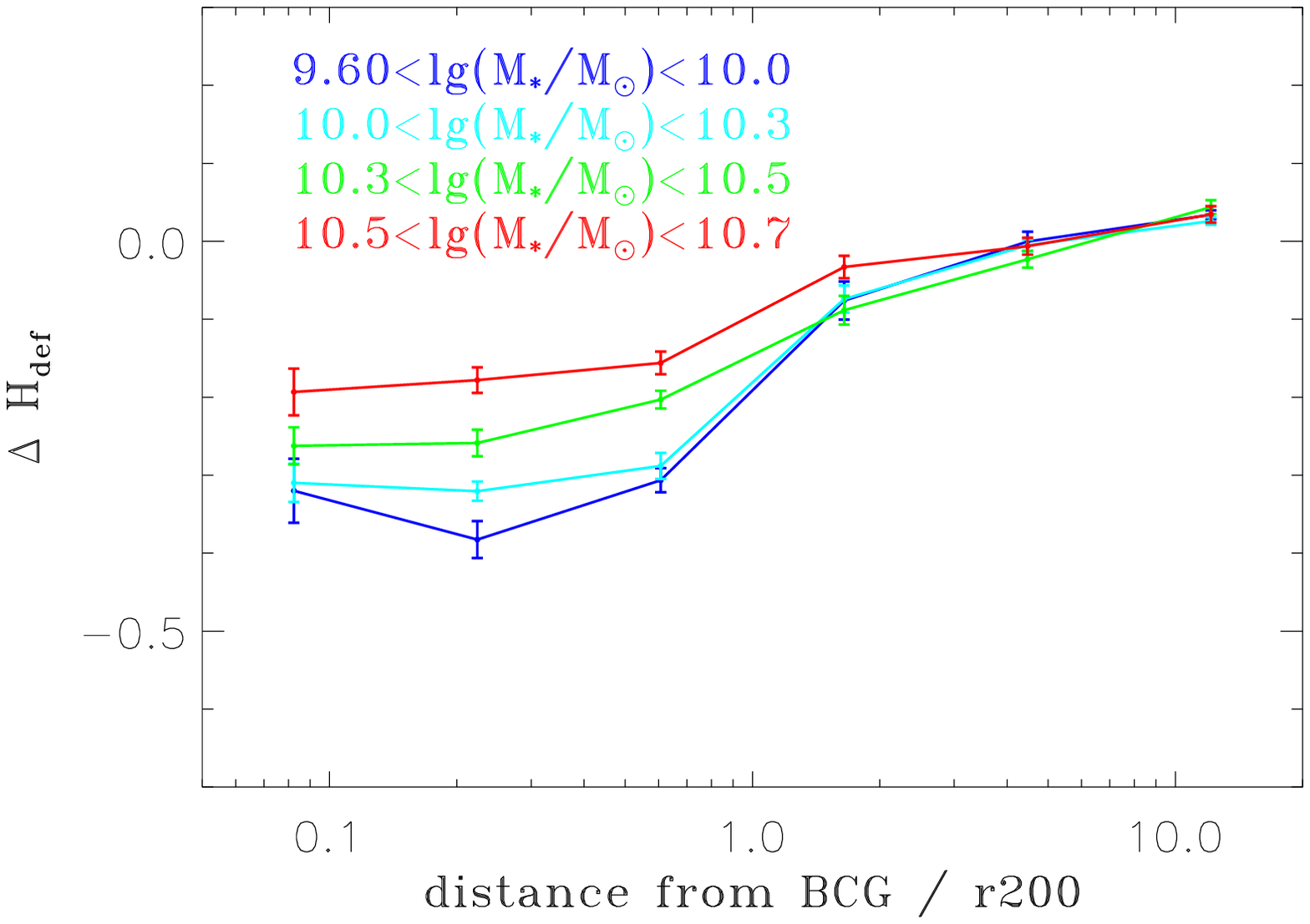,width=0.48\textwidth}
  \end{center}
\caption{Left panel: The difference in the median H{\sc i} mass fraction
with respect to galaxies of the same stellar mass in the field is   
plotted as a function of distance from the brightest cluster galaxy. Results are shown for
galaxies in different stellar mass intervals as indicated. Right panel:
same as the left panel, except that the difference in the median
H{\sc i} deficiency parameter $H_{def}$ with respect to galaxies of the same stellar mass in the field.}
\label{fig:fg_radius}
\end{figure*}

We now turn to  trends in ``inferred'' H{\sc i}
mass fraction with cluster-centric radius.  In the left panel of
Figure~\ref{fig:fg_radius} we plot the difference in the  median H{\sc i} mass
fraction with respect to field galaxies of the same stellar mass, as a function
of scaled distance from the brightest cluster galaxy (BCG). In the right panel
of Figure~\ref{fig:fg_radius} we show the same plot, except that the H{\sc i}
mass fractions are normalized to field galaxies  of the same stellar mass and
stellar surface mass density, i.e.  this is a plot of the H{\sc i} deficiency
parameter defined in the previous section.  Figure~\ref{fig:fg_radius} shows
that H{\sc i} mass fractions drop by a factor of 2-3 at the centers of clusters
with respect to ``similar'' galaxies in the field.  The effects are stronger
for less massive galaxies.  At all masses, the strongest  decrease in H{\sc i}
gas fraction occurs just at the  virial radius.  Interestingly,  the decrease
in gas fraction with radius is quite shallow in the central region of the
cluster. 
In order to understand whether this is caused by possible offsets between
the BCG and the true center of the potential well, we have repeated our 
analysis, using the stellar mass-weighted center of the clusters instead of 
the BCG as their center, and found very similar results. We also note that 
recent studies of X-ray selected clusters have shown that the offset between 
the BCG and the peak in X-ray maps is typically $\sim20$ kpc 
\citep[e.g.][]{vonderLinden-12}, much smaller than
the scale ($\sim200$ kpc) where our gas fraction profiles become shallow. 
Therefore the offset from the ture center is unlikely the reason for the
shallow profiles seen from our figure.

\begin{figure*}
  \begin{center}
    \epsfig{figure=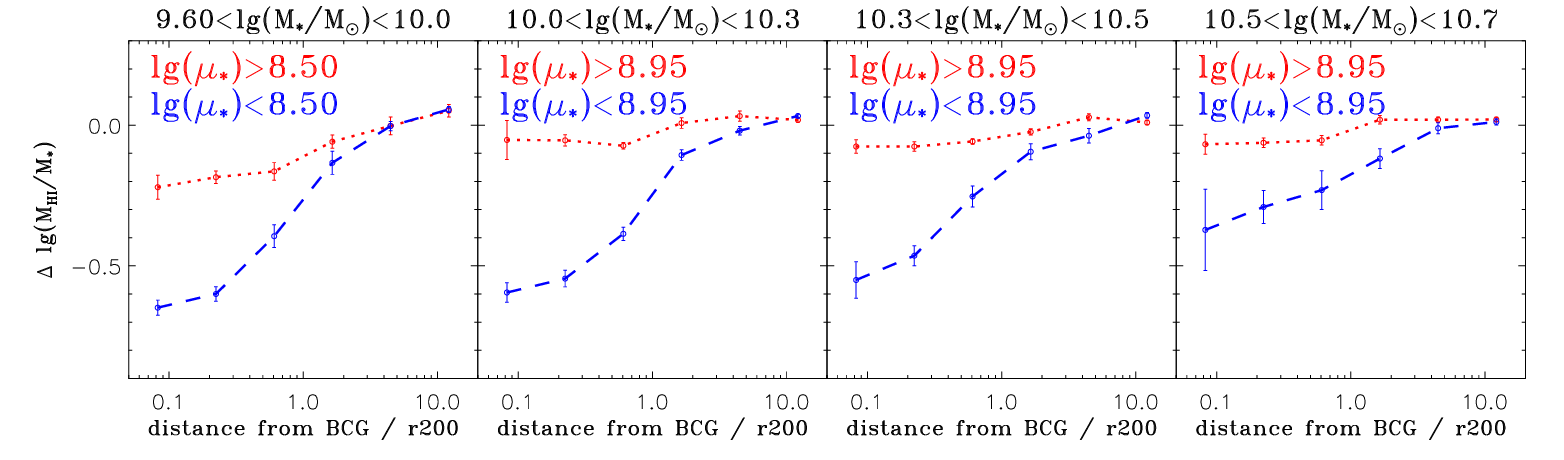,width=\textwidth}
    \epsfig{figure=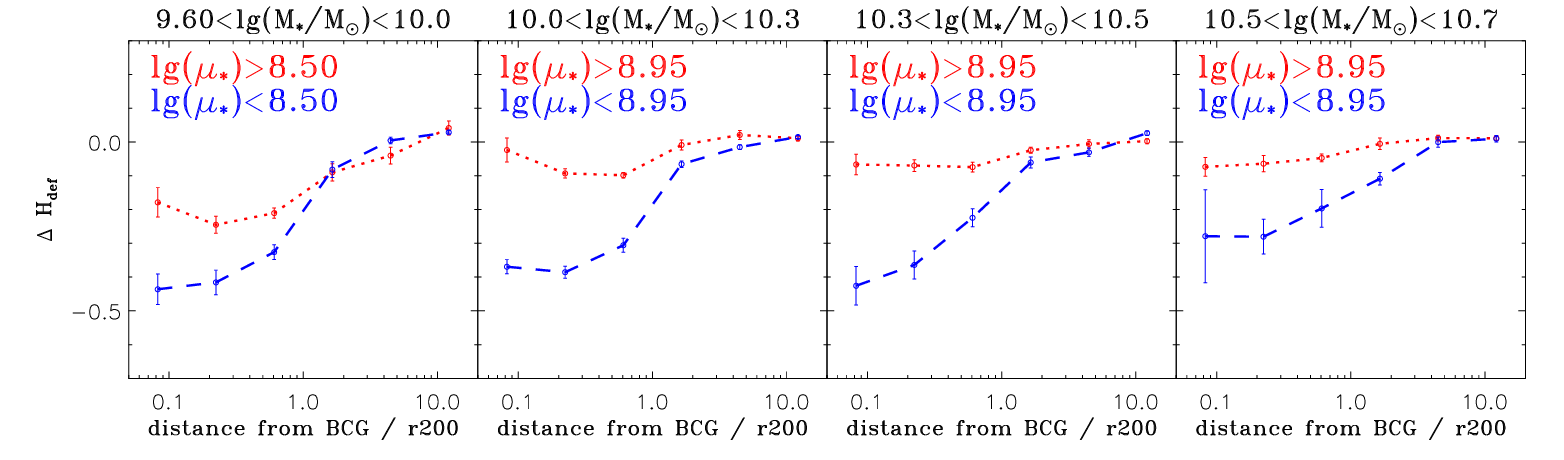,width=\textwidth}
  \end{center}
  \caption{Same as Figure~\ref{fig:fg_radius}, except for each mass interval, the galaxies have been further
separated into higher and lower $\mu_*$ subsamples, shown as red dotted and blue dashed lines.
}
\label{fig:fg_radius_mustar}
\end{figure*}

We now divide the galaxies in each stellar mass interval into two subsets with
higher and lower surface mass densities, $\log\mu_\ast$, and repeat the
analysis. The results are shown in Figure~\ref{fig:fg_radius_mustar}.  The
decrease in H{\sc i} gas content with cluster-centric distance is always
stronger for galaxies with low densities. For galaxies with stellar surface
densities greater than $10^{9} M_{\odot}$ kpc$^{-1}$, there is essentially no
change in inferred H{\sc i} mass fraction with radius.  

\begin{figure*}
 \begin{center}
   \epsfig{figure=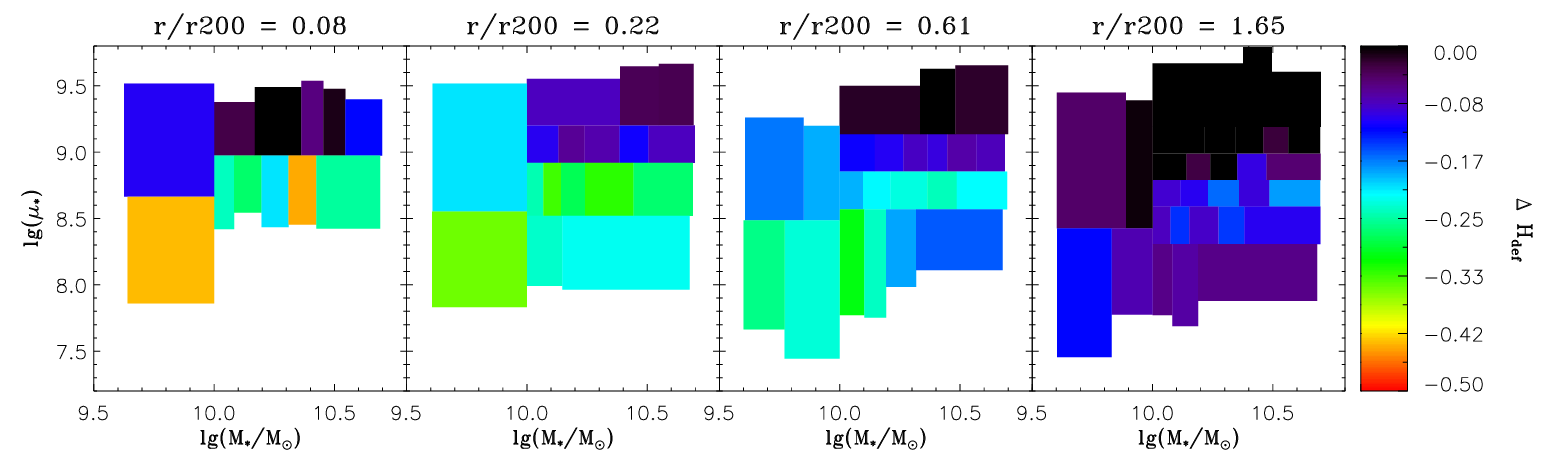,width=\textwidth}
  \end{center}
\caption{Galaxies at a given range in cluster-centric distance are binned into cells
in the 2-dimensional plane of stellar surface mass density versus stellar mass. Each
cell is colour-coded according to the mean difference in  H{\sc i} mass fraction
with respect to field galaxies of the same stellar mass and surface mass density. Results
are shown for four different ranges in cluster-centric distance as indicated on the
top of each panel.    
The cell sizes are  adjusted to enclose a fixed number of galaxies, 
whose value  are 20, 40, 80, 80, for the four panels from left to right.}
\label{fig:gas_bin}
\end{figure*}

Figure~\ref{fig:gas_bin} allows the reader to compare the dependence of  H{\sc
i} gas depletion  on stellar surface density and on stellar mass at a given
clustercentric distance.  We plot the mean difference in H{\sc i} mass fraction
with respect to field galaxies of the same stellar mass and stellar surface
mass density  in the plane of  $\mu_\ast$ versus $M_\ast$. The four panels show
results for galaxies located at different cluster-centric distances.  We used a
two-dimensional adaptive binning technique to adjust the cell size so that each
cell includes a fixed number of galaxies (20, 40, 80 and 80, respectively, for
the four cluster-centric distances).  It is clear that the decline in H{\sc i}
gas content  depends more strongly on stellar surface density than on stellar
mass.  This is true at all radii within the virial radius of the cluster, but
is especially pronounced in the first panel of Figure~\ref{fig:gas_bin}, which
shows results for galaxies in the inner cores of the clusters. 

\subsection{Dependence on halo mass}

There are a number of possible reasons why galaxies with low stellar surface
densities may lose their gas more efficiently in clusters. 

The intracluster medium may exert a drag force on the gas in galactic  disks.
If the density of the gas in the disk scales with the density of its stars,
then one might expect ram-pressure to act more effectively on galaxies with low
stellar surface densities. \footnote {We note that in practice, low density
galaxies have larger H{\sc i} gas mass fractions \citep{Catinella-10}, so this
simple assumption is unlikely to hold in detail.}  Ram-pressure scales with the
square of the velocity of the galaxy through the surrounding galaxy, so if this
is the main process at work, one would also expect to see it operate more
efficiently in clusters with higher velocity dispersions.

Alternatively, encounters between galaxies in groups and clusters may result in
stars and other material being pulled out of galaxies. Tidal stripping is most
effective when galaxies encounter each other  with relative velocities
comparable to the internal velocity dispersion of their stars.  Tidal stripping
is thus not thought to be effective in rich cluster environments, where
galaxies are moving with average velocities in excess of 1000 km/s. 

\begin{figure*}                     
  \begin{center}                   
    \epsfig{figure=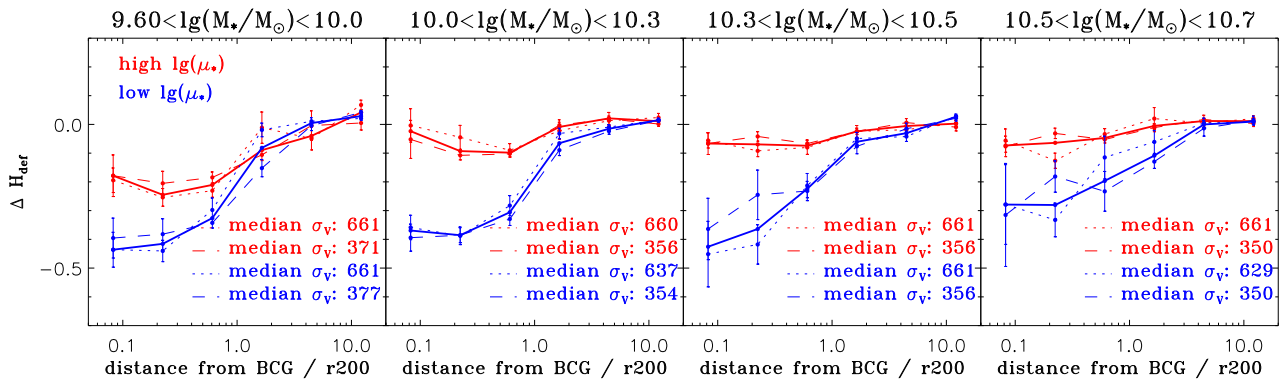,width=\textwidth}
  \end{center}                     
 \caption{Same as the bottom panel of Figure~\ref{fig:fg_radius_mustar}, except that galaxies have been divided into
two further subsamples according to the velocity dispersion of their host cluster. 
A cut at  $\sigma$= 500 $\rm km/s$ is applied and the median velocity
dispersions clusters hosting the galaxies in the various subsamples are labelled on
each panel.}
\label{fig:fg_radius_mustar_sigmav}
\end{figure*}

In Figure~\ref{fig:fg_radius_mustar_sigmav}, we divide the galaxies in each
stellar mass and stellar surface density subsample into two further subsets
according to the velocity dispersion of their associated clusters.  Our main
conclusion is that we do not find any dependence of the gas-radius relation on
cluster velocity dispersion, even for  low stellar surface mass density
galaxies. Considering the relatively large scatter in the velocity
dispersion-halo mass relation \citep[e.g.][]{Weinmann-10}, we would like 
to point out that our result does not neccessarily imply that there is no
trend in the gas-radius relation with dark matter halo mass.

\section{Summary}

In this paper, we have analyzed galaxies in 300 nearby groups and clusters
identified in the Sloan Digital Sky Survey using a  photometric gas mass
indicator that is useful for estimating the degree to which the interstellar
medium of a cluster galaxy has been depleted.  We  study the radial dependence
of inferred gas mass fractions for galaxies of different stellar masses and
stellar surface densities. Our main results may be summarized as follows. 

\begin{itemize}
\item The H{\sc i} mass fraction of galaxies decrease by a factor of $2-3$ from
the outskirts of clusters to their centres. The decrease in gas fraction
is most pronounced around the virial radius and is strongest for low mass galaxies.  

\item At fixed stellar mass and at fixed clustercentric distance, the depletion
of gas in cluster galaxies clearly depends on the stellar surface mass density
of the galaxy, in the sense that low density galaxies are more H{\sc i}
deficient. 

\item An analysis of depletion trends in the two-dimensional plane of stellar
mass $M_*$  1and stellar mass surface density $\mu_*$ reveals that gas
depletion at fixed clustercentric radius is  more sensitive to the density of a
galaxy than to its mass. 

\item The  gas mass fraction-radius relations exhibit little dependence on the
velocity dispersion of the cluster. 

\end{itemize}

We suggest that low density galaxies are more easily depleted of their gas,
because they are more easily affected by ram-pressure and/or tidal forces. 

In recent work , \citet{Fabello-12} compared the environmental density
dependence of the atomic gas mass fractions of nearby galaxies with that  of
their central and global specific star formation rates.  For galaxies less
massive than $10^{10.5}M_\odot$, the authors found both the H{\sc i} mass
fraction and the sSFR to decrease with increasing density, with the H{\sc i}
mass fraction exhibiting stronger trends than the sSFR. This was interpreted as
evidence for ram-pressure stripping of atomic gas from the outer disks of
low-mass galaxies.  The authors also  compared their results with predictions
from the semi-analytic model (SAM) of \citet{Guo-11}. They found the opposite
trend in the models: the decline in H{\sc i} mass fraction with density was
weaker than the decline in sSFR. This  indicated that the recipe of gas
stripping assumed in the current SAMs, in which only the diffuse cold gas
surrounding satellite galaxies is stripped (often called ``strangulation''), is
insufficient, and ram-pressure stripping of the cold interstellar medium is
likely to play a significant role.

\cite{Li-12b} studied the bias in the clustering of H{\sc i}-rich and H{\sc
i}-poor galaxies with respect to galaxies with normal H{\sc i} content on
scales between 100 kpc and $\sim 5$ Mpc, and compared the results with
predictions from the current semi-analytic models (SAMs) of galaxy formation of
\citet{Fu-10} and \citet{Guo-11}. They found that, for the H{\sc i}-deficient
population, the strongest bias effects arise when the H{\sc i} deficiency is
defined in comparison to galaxies of the same stellar mass and size. This is
not produced by the SAMs, where the quenching of star formation does not depend
on the internal structure of galaxies. The authors proposed that the
disagreement between the observations and the models might be resolved, if
processes such as ram-pressure stripping, which depend on the density of the
interstellar medium (ISM), are included in the models.

The results of this paper point in much the same direction. The finding that is
somewhat puzzling is that the  gas mass fraction-radius relation does not
depend significantly on the velocity dispersion of the cluster, as might be
expected if ram-pressure stripping is the main process at work in groups and
clusters. One possible solution is that {\em both} tidal and ram-pressure
stripping are at work and they act on the galaxy population in such a way as to
cancel out any significant velocity-dispersion dependent effects.  There have
been recent  detailed studies of ram-pressure stripping and tidal interactions
in individual galaxy groups with extensive  coverage of the X-ray emitting hot
gas  from $Chandra$ observations and detailed H{\sc i} maps from Very Large
Array (VLA) mosaic observations that have indicated that both ram-pressure and
tidal stripping can be at work at the same time in a single group
\citep{Rasmussen-12}. Next generation X-ray observations from the eROSITA
satellite and wide-field H{\sc i} surveys planned at the Westerbork telescope
\citep[APERTIF;][]{Verheijen-08} and at the Australian SKA Pathfinder telescope
(ASKAP) will clarify the relative importance of these processes as a function
of dark matter halo mass and location of galaxy within their halos.

\section*{Acknowledgments}

WZ and CL thank the Max-Planck  Institute for  Astrophysics (MPA)  for warm
hospitality  while  this  work  was  being completed.  It is a pleasure to
thank the anonymous referee for helpful comments.  This work is supported by
the Chinese  National Natural Science Foundation  grants 10903011 and 11173045,
and the CAS/SAFEA International Partnership Program for  Creative  Research
Teams  (KJCX2-YW-T23).  CL acknowledges the support of the 100 Talents Program
of Chinese Academy of Sciences (CAS),  Shanghai Pujiang Program (no.
11PJ1411600) and the exchange program between Max Planck Society and CAS.  GK
thank the Aspen Center for Physics and the NSF Grant \#1066293 for hospitality
during the writing of this paper.  TX acknowledges the support of the LAMOST
postdoctoral fellowship.

Funding for the SDSS and SDSS-II has been provided by the Alfred P.  Sloan
Foundation, the Participating Institutions, the National Science Foundation,
the U.S. Department of Energy, the National Aeronautics and Space
Administration, the Japanese Monbukagakusho, the Max Planck Society, and the
Higher Education Funding Council for England. The SDSS Web Site is
http://www.sdss.org/.
 
The SDSS is managed by the Astrophysical Research Consortium for the
Participating Institutions. The Participating Institutions are the American
Museum of Natural History, Astrophysical Institute Potsdam, University of
Basel, University of Cambridge, Case Western Reserve University, University of
Chicago, Drexel University, Fermilab, the Institute for Advanced Study, the
Japan Participation Group, Johns Hopkins University, the Joint Institute for
Nuclear Astrophysics, the Kavli Institute for Particle Astrophysics and
Cosmology, the Korean Scientist Group, the Chinese Academy of Sciences
(LAMOST), Los Alamos National Laboratory, the Max-Planck-Institute for
Astronomy (MPIA), the Max-Planck-Institute for Astrophysics (MPA), New Mexico
State University, Ohio State University, University of Pittsburgh, University
of Portsmouth, Princeton University, the United States Naval Observatory, and
the University of Washington.

\bibliography{ref,addon}

\label{lastpage}

\end{document}